\documentclass[conference]{IEEEtran}
\IEEEoverridecommandlockouts
\usepackage{cite}
\usepackage{amsmath,amssymb,amsfonts}
\usepackage{graphicx}
\usepackage{textcomp}
\usepackage{xcolor}
\usepackage[font=footnotesize,labelfont=bf]{caption}
\def\BibTeX{{\rm B\kern-.05em{\sc i\kern-.025em b}\kern-.08em
    T\kern-.1667em\lower.7ex\hbox{E}\kern-.125emX}}
\usepackage{setspace}

\begin{document}
\title{An Analogy of Frequency Droop Control for Grid-forming Sources}
\author{\IEEEauthorblockN{Minghui Lu, Brett Ross}
\IEEEauthorblockA{Pacific Northwest National Laboratory, WA, USA \\ Email: minghui.lu@pnnl.gov}}

\maketitle

\begin{abstract}
In this paper, we present an analogy for a power system dominated by grid-forming (GFM) sources that proves to be a powerful visualization tool for analyses of power flow, frequency regulation, and power dispatch.
Frequency droop characteristics of a typical GFM source are exactly reflected by an ordinary model of water vessels.
The frequency is represented by visible water levels while the droop slope is reified by the vessel sizes.
This proposed analogy allows us to use the intuitive water-flow phenomenon to explain the abstract power-flow problems.
The grid integration of renewables via GFM inverters is interestingly simulated by a vessel connected to an infinite water tank.
This paper also provides a means for demonstrating issues to audiences with little or no background in power systems.
Finally, the proposal is verified by simulation results.
\end{abstract}


\begin{figure}[b]
\centering
\includegraphics[width = 0.48\textwidth]{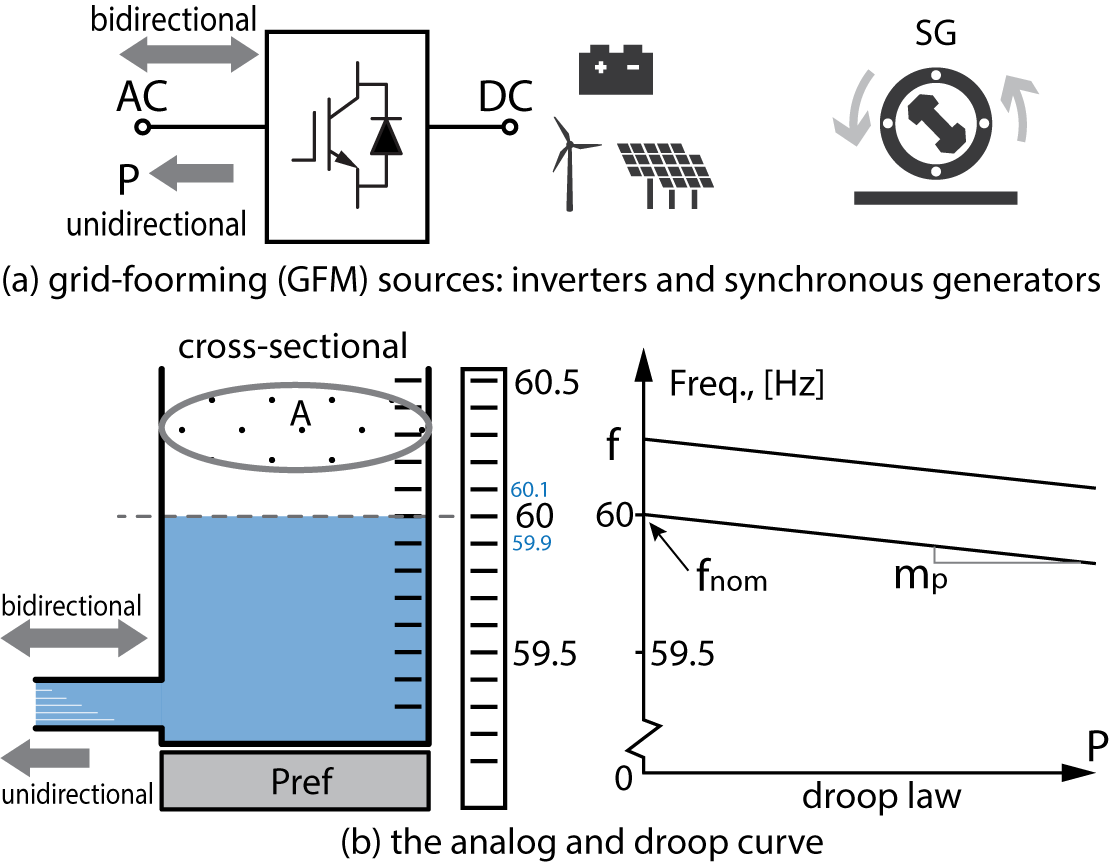}
\caption{Grid-forming sources and the associated analogy.}
\label{fig:inv}
\end{figure}
\section{Introduction}
The integration of the inverter-based resources (IBRs) poses challenges in maintaining the stability, frequency, and voltage regulation of the grid.
One of the key techniques used for integrating IBRs into grid is the grid-forming (GFM) control.
GFM sources rely on actively adjusting frequency in response to changes in real powers in order to achieve synchronization and reach frequency equilibrium with other GFM sources in the associated grid~\cite{Roadmap2020}.
Ordinary GFM controls have largely been based on droop-control laws that draw inspiration from the quasi-steady-state operation of synchronous generators (SGs)~\cite{Chandorkar1993}.
Along similar lines, virtual synchronous machine methods are focused on direct emulation of machine dynamics~\cite{Zhong2011, Liu2017}.
Departing from machine-inspired approaches, dispatchable virtual oscillator control (dVOC) is a recent control strategy where inverters are programmed to emulate the dynamics of nonlinear limit-cycle oscillators~\cite{Lu2019, Seo2019, Lu2022}.

For all GFM controls above, the typical droop characteristics, as shown in Fig.~\ref{fig:inv}, are the core parts of the controllers, inverters regulate their output frequencies and voltages based on the real and reactive powers.
In particular, the linear frequency droop curve is given in Fig.~\ref{fig:inv}(b).
Carefully chosen analogies are essential to gain understanding of the physical world and this in turn is essential in order to communicate difficult concepts.
Giving a proper explanation of the roles that grid-forming inverters play in an electric power system as well as the technical concepts of droop control, frequency regulation, and power dispatch requires more than a few mathematical equations; accordingly, this article will attempt to explain many of these concepts using analogies.
By comparing GFM sources and the communicating vessels, this paper shows that a close analogy exists between the two.
This is probably a lot to take in when reading it as prose, so Table~\ref{tab1} is a table to summarize.
It demonstrates that the frequency droop behavior of interconnected GFM sources well analogized by water flow among communicating vessels.

The main contributions of this paper can be summarized as follows: We 1) establish the analogy model of GFM sources to achieve the deeper understanding of technical concepts (e.g., frequency regulation and power dispatch); 2) This work provides visualization for the frequency control of droop-controlled sources.

\begin{table}[b]
\centering
\begin{tabular}{ l|p{3.5cm}|p{3cm} } 
\hline \hline
 & Communicating vessels & GFM sources \\ 
\hline \hline
$f$ & water level & actual frequency \\
\(f_\mathrm{nom}\) & base reference level & nominal frequency \\
$A$ & cross-sectional area & reciprocal of $m_\mathrm{p}$ \\
$m_\mathrm{p}$ & reciprocal of $A$ & droop slope \\
$P$ & outflow water amount & real power \\
$P_\mathrm{ref}$ & block elevation & power setpoint \\
$L$ & pipes & electric cables \\
\hline
\end{tabular}
\captionof{table}{The table comparing communicating vessels and GFM sources.}\label{tab1}
\end{table}

The rest of this paper is organized as follows: The overall system and research motivation are introduced in Section~\ref{sec:idea}.
The basic idea, principle, and implementation of this proposal are given in Section~\ref{sec:results}.
Finally, concluding and future-work statements are in Section~\ref{sec:conclusion}.

\section{System Description}\label{sec:idea}
\subsection{Grid-forming (GFM) Controls}
Figure~\ref{fig:inv}(a) shows typical GFM sources that can be either inverters or synchronous generators.
For inverters, the power-flow direction could be bidirectional or unidirectional depending on the types of dc-side sources.
As shown in Fig.~\ref{fig:inv}(b), the GFM sources rely on actively adjusting frequency, \(f\), in response to changes in real powers, \(P\), in order to achieve synchronization with other GFM sources in the associated grid.
The control goal is to maintain frequency and real power within safe boundaries around the scheduled value at all times~\cite{Eto2010}.

Despite control varieties, the classic frequency droop law can be expressed as
\begin{equation}
f = f_\mathrm{nom} - m_\mathrm{p} (P - P_\mathrm{ref}),
\label{eq:droop}
\end{equation}
where $f$ is actual frequency, $f_\mathrm{nom}$ is nominal frequency, $m_\mathrm{p}$ is the droop slope, and $P_\mathrm{ref}$ is the real-power setpoint.
In general, the value of droop slope can be selected as 
\begin{equation}
m_\mathrm{p} = \frac{\Delta f}{S_\mathrm{rated}},
\label{eq:slope}
\end{equation}
where the \(\Delta f\) is desired frequency drop and \(S_\mathrm{rated}\) is rated power.
In addition to the linear frequency control given in~\eqref{eq:droop}, publications also report nonlinear droop varieties to achieve specific goals~\cite{Chen2017}.
We will have further discussion on nonlinear controls and their analogy later.

\subsection{The Analog}
\begin{figure}[b]
\centering
\includegraphics[width = 0.49\textwidth]{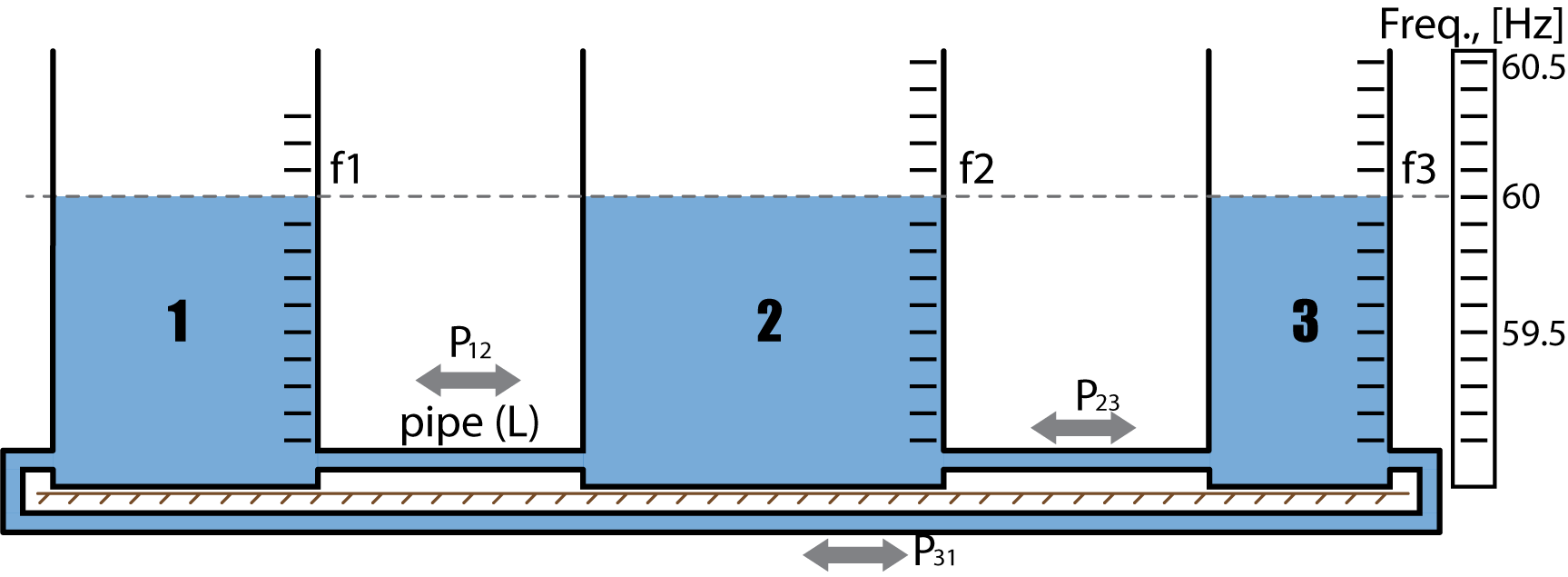}
\caption{The communicating-vessel analogy for frequency regulation of interconnected GFM sources.}
\label{fig:analogy}
\end{figure}

Consider a system of communicating vessels shown in Fig.~\ref{fig:analogy}, three vessels with various sizes are placed on the same horizontal surface, their bottoms are interconnected via pipes.
We can imagine a scenario, if we add water to vessel 2, the water level will be temporarily higher than that in vessels 1 and 3.
Due to the difference in hydraulic pressure, some amount of water will flow through the pipes into vessels 1 and 3, the water levels in all the vessels will eventually settle at a same new level in steady state.
This phenomenon inspired us to associate the water level with the frequency in ac power systems, where the frequencies of all interconnected GFM sources will be equal in the quasi-steady state.

Therefore, we are motivated to make a side-by-side analogy, in which the vessels exactly mirror the GFM sources, see Fig.~\ref{fig:inv}(b) and Fig.~\ref{fig:analogy}.
The water level in each vessel reflects the actual output voltage frequency, \(f_k, (k = 1, 2, 3)\), whose nominal value is 60 Hz.
The ruler on the right is the frequency indicator, measuring the water levels in real time.
As Fig.~\ref{fig:inv}(b), if the water outflows, the frequency will linearly decrease with the water amount; Similarly, if the water inflows, the frequency will linearly increase with the water amount.
This intuitively reflects the frequency droop feature of the whole GFM system.
The pipes connecting the vessels represent the transmission cables that carry directional power flows.
To summarize, Tab.~\ref{tab1} gives a full map of the side-by-side comparison.

\subsection{Droop Slope and Proportional Power Sharing}
Although the steady-state water levels of vessels are equal, the water amount increase in each vessel is not always identical.
In fact, the change of water amount is proportional to the vessel sizes or diameters, a larger vessel (the vessel 2 in Fig.~\ref{fig:analogy}) will allocate more amount increase whereas a smaller vessel (the vessel 3) will share less amount increase.
This actually reflects the proportional sharing capabilities of interconnected GFM sources~\cite{Roadmap2020}.

Technically, the droop slope of \(P-f\) curve, \(m_\mathrm{p}\), is inversely proportional to the cross-sectional area, $A$, or
\begin{equation}
m_\mathrm{p} = \frac{1}{A}.
\end{equation}
It makes sense that a GFM source with larger rated power or capacity should have a smaller droop slope to achieve proportional power sharing, which also corresponds to the droop slope definition in~\eqref{eq:slope}.
This characteristic allows multiple inverters to share load proportionally to their rated powers.

\begin{figure}[t]
\centering
\includegraphics[width=0.49\textwidth]{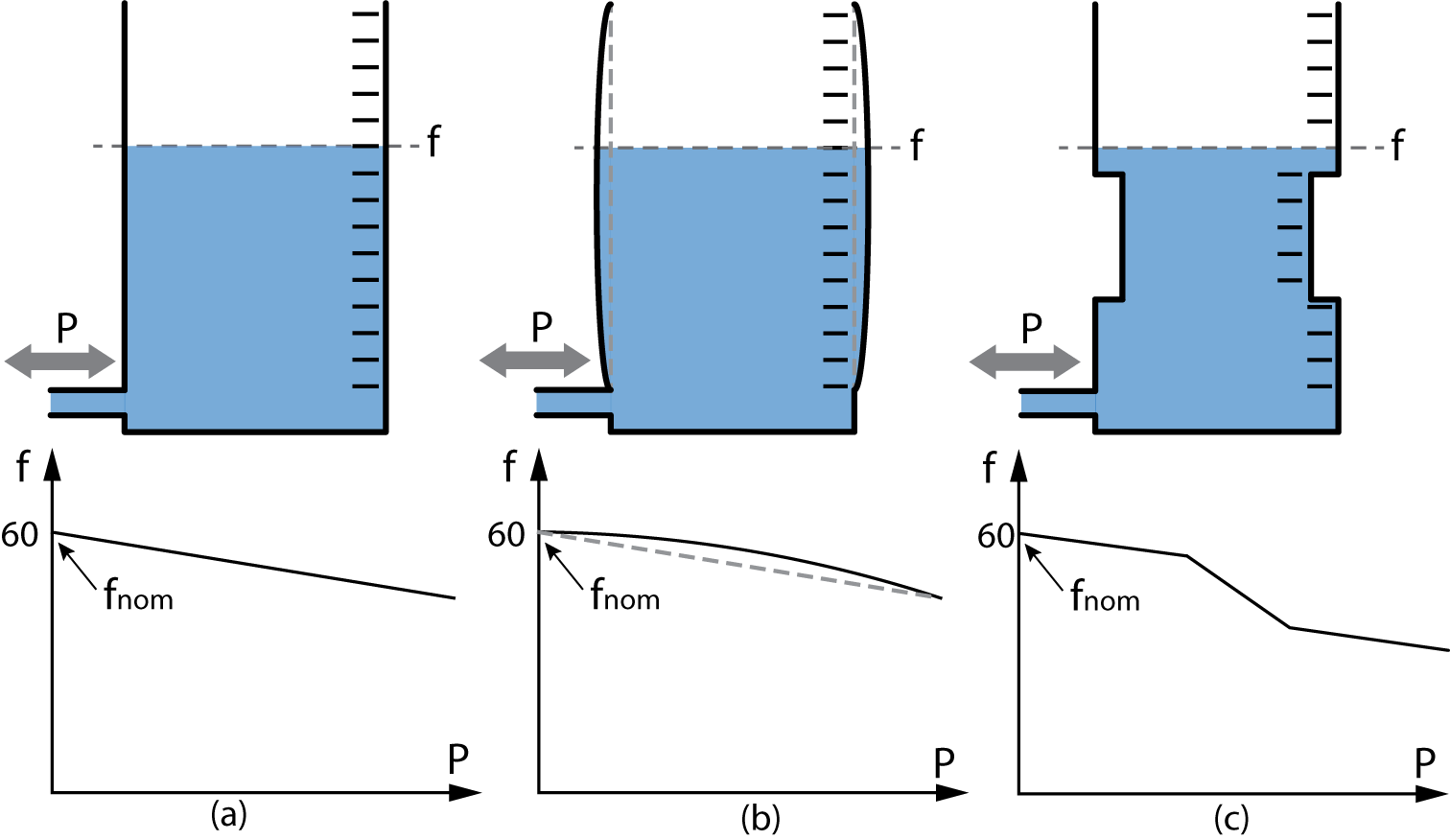}
\caption{Two nonlinear droop controls and their analog models, (a) linear, (b) convex curved, (c) piecewise linear.}
\label{fig:nonlinear}
\end{figure}
\subsection{Nonlinear Droop Controls}
Although the linear droop control is relatively simple, robust, and efficient, it may not guarantee the system operation at the most economic point due to the nonlinear operating cost behaviors~\cite{Abdelgabir2022, Fatih2017}.
Chen et al.~\cite{Chen2017} proposed a cost-based droop control scheme for islanded microgrids considering the optimization of the total generation costs.
\cite{Nutkani2014} also reported a similar droop control strategy with consideration of operating costs.
Different from linear droop controls, this type of nonlinear schemes exhibit an obviously convex curve as shown in Fig.~\ref{fig:nonlinear}(a).
It shows an intuitive representation by a vessel with a slightly curved body, which indicates the droop slope, \(m_\mathrm{p}\), is obviously nonlinear.

Another form of nonlinear droop controls owes the piecewise linear feature~\cite{Igor2020}, which can be expediently displayed by a particular vessel in Fig.~\ref{fig:nonlinear}(b).
Theoretically, the proposed analogy can exhibit any nonlinear droop control by a vessel with the desired and customized body shape, will greatly facilitate to visualize the frequency regulation of nonlinear GFM controls.



\section{Analogy Applications}
This section will continue to explore the potential of applying the proposed analogy to real-world applications.

\begin{figure}[t]
\centering
\includegraphics[width = 0.5\textwidth]{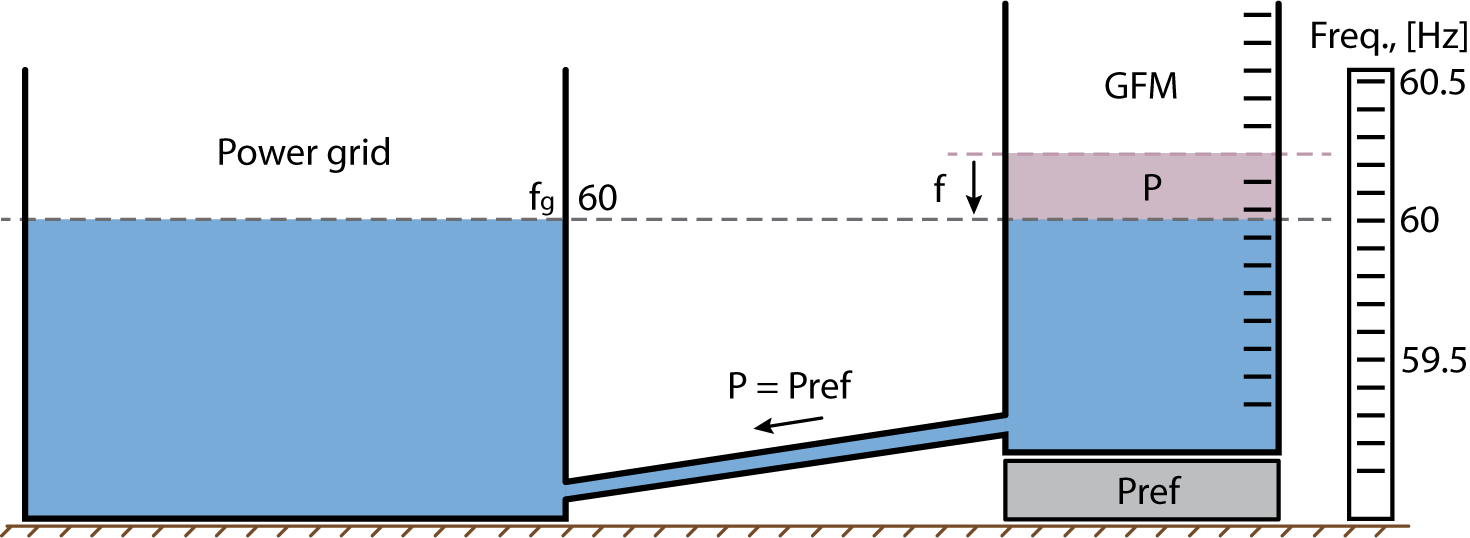}
\caption{Analogy of connecting a GFM inverter to the bulk power grid.}
\label{fig:grid_connection}
\end{figure}

\subsection{Power Setpoints and Grid Connection}
We start with the grid connection of a single GFM source, the grid is regarded as a stiff voltage source with infinite inertia, its frequency is fixed at \(f_\mathrm{g} = 60.0\) Hz.
In the analogy, Fig.~\ref{fig:grid_connection} describes the power grid as a water tank with infinite water volume, its water level is fixed though some water is added or removed from the tank.
A vessel is connected to the water tank via a pipe, which mirrors the grid-connection of a GFM source.
The height at which the vessel is placed means the real-power setpoint, \(P_\mathrm{ref}\), which determines the amount of water flowing into or from the infinite water tank (denoted as $P$ in Fig.~\ref{fig:grid_connection}).
In fact, they are equal to each other, i.e., $P = P_\mathrm{ref}$, which means it achieves the target power tracking.

\begin{figure}[b]
\centering
\includegraphics[width = 0.5\textwidth]{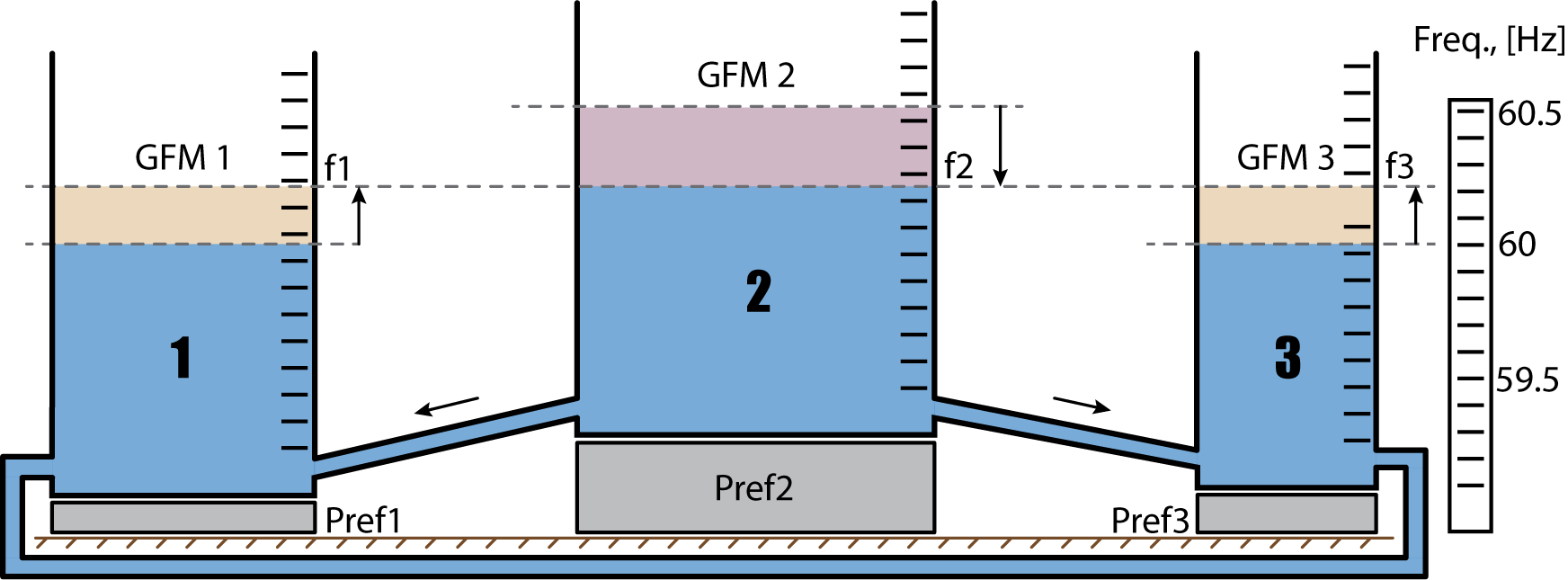}
\caption{Analogy of interconnected GFM sources.}
\label{fig:power_dispatch}
\end{figure}
\subsection{Interconnected GFM Sources}
Next, we extend the analogy to the application of interconnected GFM sources, which reflects the core part of an electric power system. 
Figure~\ref{fig:power_dispatch} shows three vessels lifted to the horizontal surfaces at different levels, which means each GFM source is power-dispatchable and has its own real-power setpoint, \(P_{\mathrm{ref},k}, (k = 1, 2, 3)\).
They are coupled together (via pipes) just as they are in a power system via transmission lines and transformers, etc;
they collectively maintain the water level just as GFM generators and/or inverters jointly maintain the power system frequency.
We can actually kind of visualize that the slope \(m_{\mathrm{p},k}\) and power setpoint \(P_{\mathrm{ref},k}\) are key parameters during power system operations.
Together, they determine how much water is allocated to each vessel, i.e., how much power is dispatched by each GFM source.

In a sense, it is viable to alter the droop slope, \(m_{\mathrm{p},k}\), to achieve the target output power, but the slope usually has been already pre-designed as specifications.
Arbitrarily changing it might cause undesired outcomes, for example system malfunctions and instability. 
Therefore, the common approach is to shift the power setpoints \(P_{\mathrm{ref},k}\) up/down to achieve the target power, which is similar to the dispatch in grid-connected mode.
Note that the key difference with grid-connected operations is that it always refers to a particular baseline, the grid frequency, whereas the power dispatch in Fig.~\ref{fig:power_dispatch} refers to a new frequency closed to 60 Hz but determined by all three GFM sources.

\begin{figure}[b]
\centering
\includegraphics[width=0.5\textwidth]{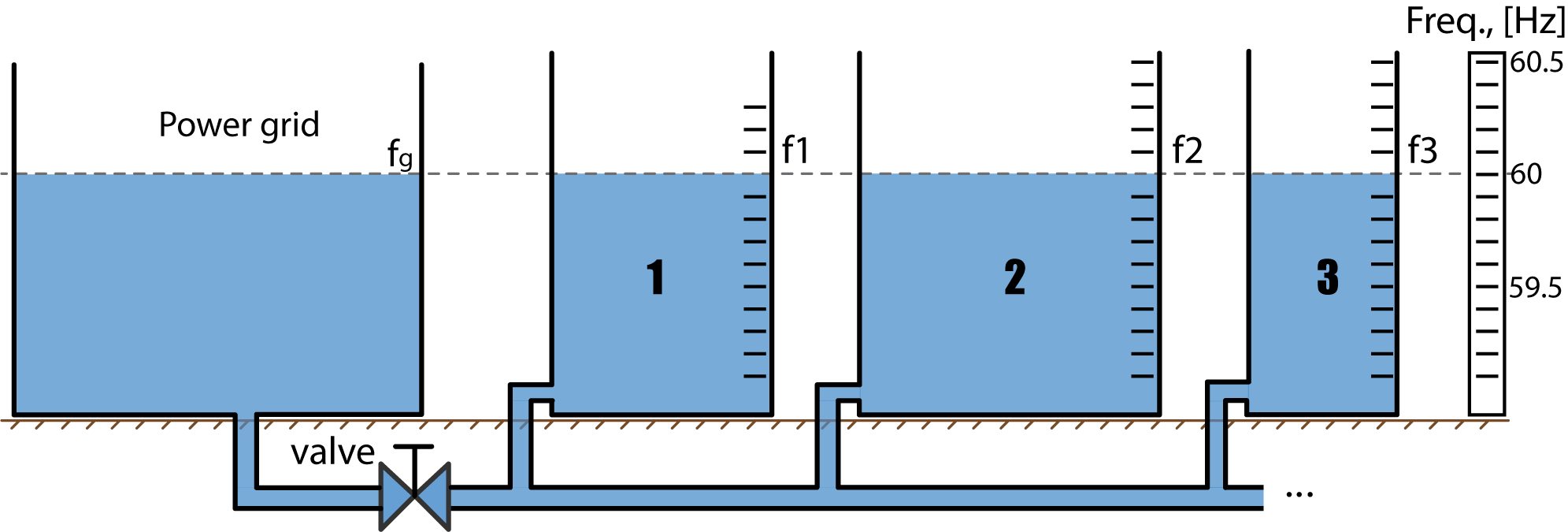}
\caption{Analogy of Microgrid in both islanded and grid-connected modes depending on the status of the valve.}
\label{fig:microgrid}
\end{figure}
\subsection{Microgrid}
The analogy can also be applicable to the microgrid that is a small-scale, localized electrical grid that can operate independently or in connection to a larger electric grid.
Figure~\ref{fig:microgrid} shows that all GFM sources are connected to the Point of Common Coupling (PCC)~\cite{Joan2012}.
The status of the valve determine the operating mode of such a system, if it is open, the system operates in grid-connected mode; if the valve is closed, the system is islanded, all three GFM sources operate based on the frequency droop law.

Although the steady-state water levels in the vessels are equal, the increase in water volume for each vessel is not necessarily the same.
In fact, the change of water amount is proportional to the vessel sizes or diameters, a larger vessel (the vessel 2 in Fig.~\ref{fig:analogy}) will allocate more amount increase whereas a smaller vessel (the vessel 3) will share less amount increase.
This actually reflects the proportional sharing capabilities of interconnected GFM sources.

\section{Demonstration}\label{sec:results}
We now illustrate the analogy through simulation results for grid-connected, interconnected, and microgrid modes of operation.

\subsection{Grid-Connected GFM Source}
\begin{figure}[t]
\centering \vspace{-15pt}
\includegraphics[width = 0.48\textwidth]{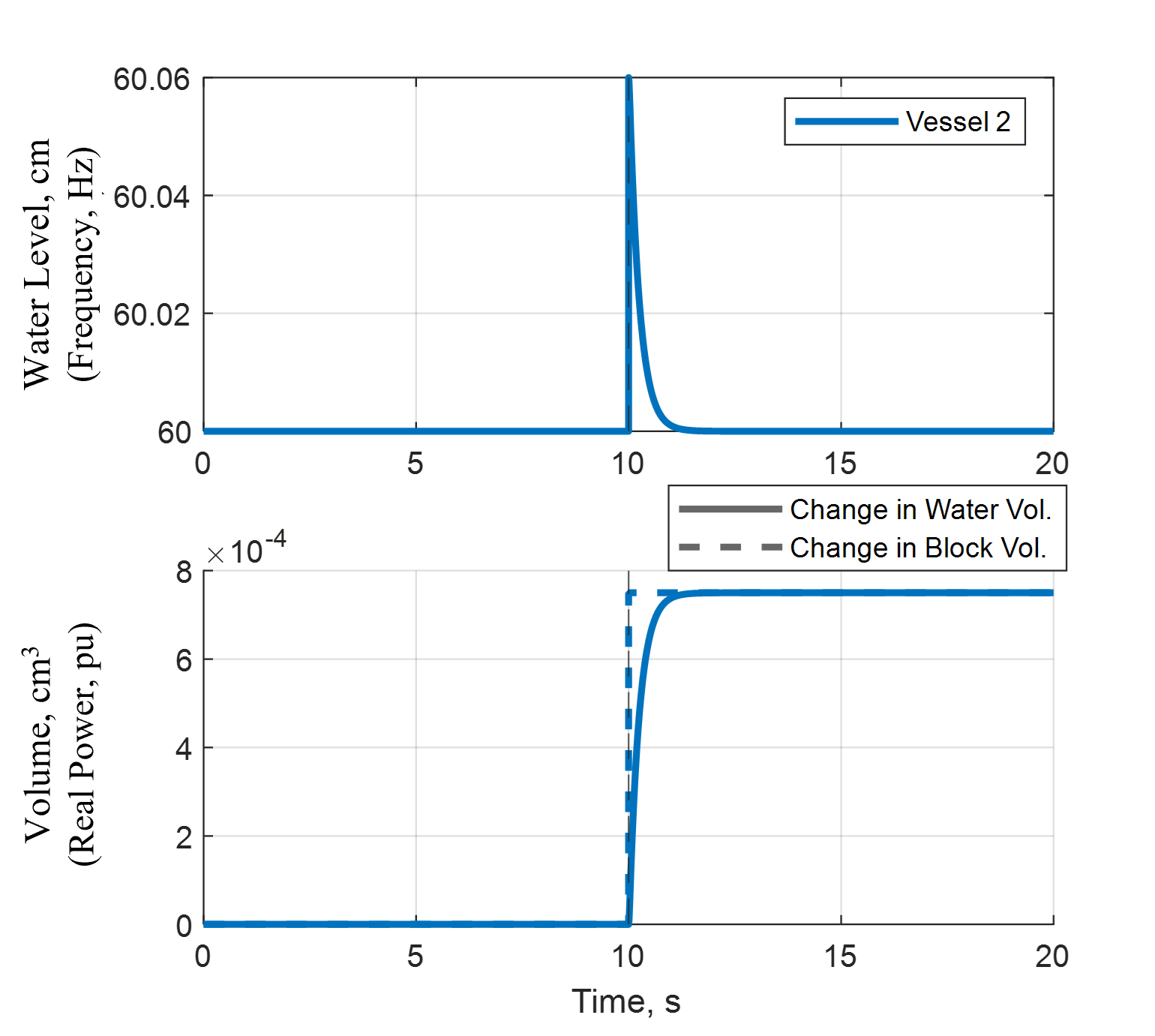}
\caption{Demonstration of connecting a GFM inverter to the bulk power grid.}
\label{fig:demo_grid}
\end{figure}

Figure~\ref{fig:demo_grid} illustrates the response of a single vessel (vessel 2) connected to an infinite vessel as introduced in Fig.~\ref{fig:grid_connection}.
At t = 0, the vessel is at the same level as the infinite vessel and the block labeled ``\(P_\mathrm{ref}\)" is absent (i.e., \(P_\mathrm{ref}\) = 0).
At t = 10 s, the height of vessel 2 is increased by placing it on the block (e.g., \(P_\mathrm{ref}\) is set to a positive constant).
This brings the total height of the water in vessel 2 to above that of the infinite vessel, and water flows into the infinite vessel.
Because the water level in the infinite vessel does not rise at all, the total volume of water that leaves vessel 2 is exactly equal to the volume of the vessel (we assume that the block has the same cross-section as the vessel).
This behavior is analogous to the situation where a GFM source connects to a bulk grid and achieves real power tracking without any steady-state error.
Note that, in the bottom plot, volume leaving the vessel (as opposed to entering) is given a positive sign convention.
This is done to make it easier to visualize how this volume change relates to real power.

\subsection{Interconnected GFM Sources}
\begin{figure}
\centering\vspace{-20pt}
\includegraphics[width = 0.49\textwidth]{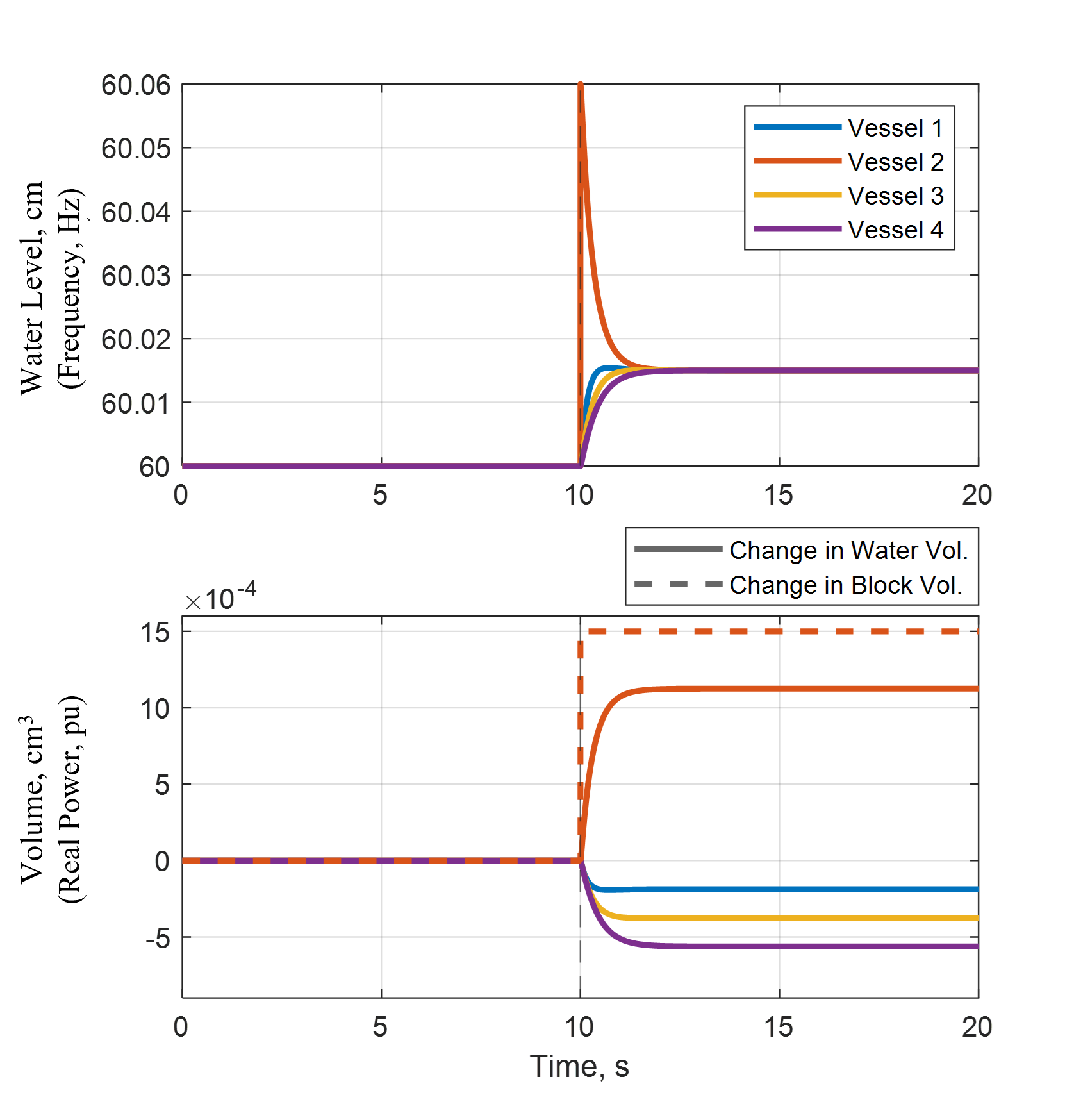}
\vspace{-5pt}
\caption{Demonstration of interconnected GFM sources.}
\label{fig:demo_inter}
\end{figure}

Next, we demonstrate the behavior of numerous interconnected vessels, which correspond to interconnected GFM sources as illustrated back in Fig.~\ref{fig:power_dispatch}.
This is done by simulating the behavior of a four-vessel system, where all the vessels are of similar sizes and all the interconnecting pipes are of similar lengths and cross-sectional areas. 
Figure~\ref{fig:demo_inter} illustrates the response of this four-vessel system to an increase in the height of vessel 2 at t = 10 s. Initially, all vessels are at the same height (i.e., \(P_\mathrm{ref}\) = 0 for all vessels) and all the water heights are the same (60 cm).
When the height of vessel 2 is increased, water begins to flow out of vessel 2 and into the other three vessels (the exiting water volume for vessel 2 is positive and negative for the other vessels). 
Vessels 1, 2, 3, and 4 have cross sectional areas of 1.25, 2.5, 2.5, and 3.75 cm\textsuperscript{2}, respectively.
This determines their relative change in volumes, which is analogous to the power sharing behavior of droop control.
Vessel 1 has the smallest area, which corresponds to the largest droop gain and thus the lowest participation in sharing the distribution of water. 
Vessels 2 and 3 have equal areas and share equal amounts of the changing water volume\textemdash{}the difference between vessel 2's set change in volume (the dashed orange line) and the actual change in volume (the solid orange line) is roughly the same as the difference between vessel 3's set change in volume (zero) and its actual change in volume.
Vessel 4 has the largest cross-section and handles the most of the change in water volume.
Notice that this sharing of water prevents the volume of water in vessel 2 from changing to match the volume of the block, as was the case back in Fig.~\ref{fig:demo_grid}.
This is analogous to the error in power setpoint that GFM sources exhibit when they participate in frequency regulation via droop control.  

\subsection{GFM Sources in a Microgrid}
\begin{figure}
\centering\vspace{-15pt}
\includegraphics[width = 0.49\textwidth]{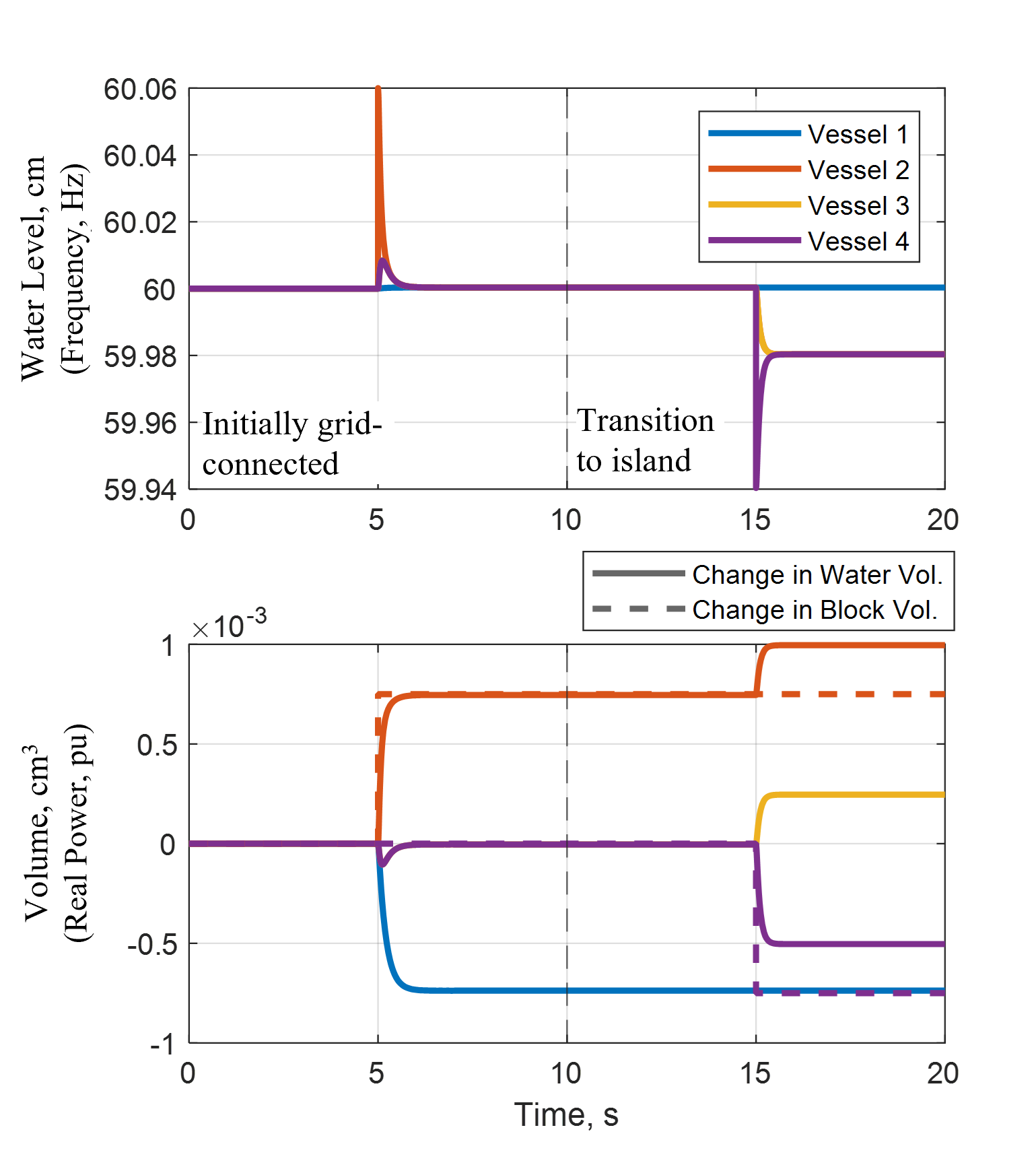}
\caption{Demonstration of GFM sources (Vessels 2, 3, and 4) connected to bulk power grid (Vessel 1).}
\label{fig:demo_micro}
\end{figure}

In our third example, we demonstrate the behavior of GFM sources in a microgrid (refer to Fig.~\ref{fig:microgrid}) using another four-vessel system.
In this case, however, vessel 1 is altered to more represent an infinite vessel. The cross section of vessel 1, as well as that of the pipes connecting it to the other vessels, are made roughly an order-of-magnitude larger than that of the remaining three vessels. Vessels 2, 3, and 4 are used here to represent a microgrid and are all given the same cross-sectional area. Initially, all four vessels are interconnected, at the same elevation, and at the same water level (60 cm).
At t = 5 s, vessel 2 is placed on a block (Fig.~\ref{fig:demo_micro}).
Because vessel 1 is essentially an infinite vessel, vessel 2 behaves much as it did back in Fig.~\ref{fig:demo_grid}; the change in vessel 2's volume is roughly equal to the volume of the block it was placed on, and the overall water levels in the system are barely affected.
At t = 10 s, the pipes connecting vessel 1 to the other three vessels are opened\textemdash{}this represents the microgrid's transition from grid-connected to islanded mode.
At this time, there is no change in water levels. This is because there is no water flow in the pipes and thus closing valves does not change the pipe flows.
At t = 15 s, vessel 4's elevation is decreased. Physically, this can be thought of as placing it on a lowered platform or in a hole instead of on top of a block.
The net volume change is the same as was imposed on vessel 2 back at t = 5 s.
However, the overall change in the water levels of vessels 2, 3, and 4 are much more pronounced.
Vessel 4's water volume does not increase to fully replace the volume lost from the height reduction, and all three vessels share this deficit roughly equally.
Vessel 1's condition does not change at all because it is no longer connected to the other three vessels. This scenario is analogous to a sudden increase in load in an islanded microgrid.
Without the infinite vessel to support them, the smaller vessels that make up the microgrid must give up some of their volume in order to maintain the water level.
Sources in microgrids are more affected by sudden changes in load during islanded mode than in grid-connected mode, this is because they must give allocate portion of their capacity to regulating frequency via droop control. 

\section{Conclusion}\label{sec:conclusion}
This paper presents an analogy for a power system dominated by GFM sources that proves to be a powerful visualization tool for studying power flow, frequency regulation, and power dispatch.
The frequency droop characteristics of a single GFM source are exactly reflected by an accessible model of water vessel.
Moreover, this proposed analogy is verified by a set of simulation results.
The present analogy can be fruitful for the future modeling works performed by electrical engineering scientists and engineers.

\section*{Acknowledgment}
The authors would like to sincerely thank the supports from the Energy Systems Resilience Group under Pacific Northwest National Laboratory.

\footnotesize
\begin{spacing}{1}
\bibliographystyle{ieeetr}
\addcontentsline{toc}{section}{\refname}
\bibliography{bibliography}
\end{spacing}

\end{document}